\documentclass[a4paper,10pt,eqno]{article}
\usepackage{theorem}
\usepackage{latexsym,amssymb,amsfonts,amsmath}
\usepackage{graphicx}

\newtheorem{thm}{Theorem}[section]
\newtheorem{lemm}[thm]{Lemma}

\newtheorem{exa}[thm]{Example}

\newtheorem{rem}[thm]{Remark}

\theoremheaderfont{\scshape}

\newcommand{\xvec}{\ensuremath{\boldsymbol{x}}}

\newcommand{\evec}{\ensuremath{\boldsymbol{e}}}

\newcommand{\cvec}{\ensuremath{\boldsymbol{c}}}
\newcommand{\avec}{\ensuremath{\boldsymbol{a}}}
\newcommand{\uvec}{\ensuremath{\boldsymbol{u}}}

\title{{\Large {\bf STATIONARY AMPLITUDES OF QUANTUM WALKS ON THE HIGHER-DIMENSIONAL INTEGER LATTICE}}}
\author{
{\small Takashi Komatsu\footnote{komatsu-takashi-fn@ynu.ac.jp (e-mail of the corresponding author)},\qquad Norio Konno\footnote{konno-norio-bt@ynu.ac.jp}}\\
{\scriptsize  Department of Applied Mathematics, Faculty of Engineering, Yokohama National University}\\
{\scriptsize \footnotesize\it 79-5 Tokiwadai, Hodogaya, Yokohama, 240-8501, Japan}\\
}
\date{\empty}
\begin{document}
\maketitle

\par\noindent
\begin{small}
\par\noindent
{\bf Abstract}. Stationary measures of quantum walks on the one-dimensional integer lattice are  well studied. However, the stationary measure for the higher dimensional case has not been clarified. In this paper, we give the stationary amplitude for quantum walks on the $d$-dimensional integer lattice with a finite support by solving the corresponding eigenvalue problem. As a corollary, we can obtain the stationary measures of the Grover walks. In fact, the amplitude for the stationary measure is an eigenfunction with eigenvalue $1$.

\footnote[0]{
{\it Abbr. title:} Stationary amplitudes of quantum walks on the higher-dimensional integer lattice
}
\end{small}

\setcounter{equation}{0}

\section{Introduction \label{intro}}
The notion of discrete time quantum walks was introduced by Aharonov et al. \cite{adz} as a quantum counterpart of the classical random walks. Recently, the quantum walk is intensively studied in quantum physics and quantum computing \cite{mw}, \cite{por}.

The behavior of the quantum walk is quite different from that of classical random walk, e.g., ballistic spreading and localization. As for the ballistic spreading, weak limit theorems of rescaled quantum walks are reported in \cite{fgmb}, \cite{hkss}, \cite{k}, \cite{ko1}, \cite{mckb}, \cite{wkkk}. On the other hand, localization of the quantum walk is given in \cite{hkss}, \cite{ikk}, \cite{iks}, \cite{tate}, \cite{wkkk}. 

Stationary measures of quantum walks in one dimension are well studied. However, the stationary measure for the higher-dimensional case has not been clarified. Therefore, one of our basic interests is to obtain stationary measures of quantum walks on the higher-dimensional integer lattice. 

As for the stationary measure of the quantum walk in the one-dimensional lattice, the following results are reported. Konno \cite{ko2} obtained stationary measures of the three-state Grover walk. Wang et al. \cite{wlw} investigated stationary measures of the three-state Grover walk with one defect at the origin. Furthermore, Endo et al. \cite{ekk2} clarified a relation between stationary and limit measures of the three-state Grover walk. In our previous work \cite{kkk}, we investigated the stationary measures for the three-state quantum walks including the Fourier and Grover walks by solving the corresponding eigenvalue problem.
Then we found the stationary measure with a periodicity. Moreover, we could apply this situation to the three-state Fourier walk on a cycle. Stationary measures for other quantum walk models in one dimension are also studied in \cite{eekst}, \cite{ekk1}, \cite{ek}, \cite{eko}, \cite{ekst}, \cite{kls}.

Compared with the one-dimensional case, the study on the stationary measure of the quantum walk on higher-dimensional lattice is almost not known. Konno and Takei \cite{kt} showed that the set of uniform measures is contained in the set of stationary measures in more general graphs including the higher-dimensional lattice. Thus our purpose is to find the non-trivial stationary measures of the higher-dimensional integer lattice. In this paper, we present the stationary amplitude of the higher-dimensional quantum walk with a finite support by solving the corresponding eigenvalue problem. As a corollary, we obtain the stationary measures of the Grover walk in higher dimensions. Our results would be useful for quantum information processing.

This paper is organized as follows. Section \ref{quantumzddef} is devoted to the definition of discrete time quantum walks on the $d$-dimensional integer lattice. In Section \ref{zdstationary1}, we obtain stationary amplitudes and stationary measures of the Grover walks on the high-dimensional integer lattice by solving the eigenvalue problem. In Section \ref{examples}, we consider typical examples on two-and three-dimensional cases. Conclusion is given in  Section \ref{conclusion}.

\section{Discrete time quantum walks on $\mathbb{Z}^d$}\label{quantumzddef}
In this section, we give the definition of $2d$-state discrete time quantum walks on $\mathbb{Z}^d$, where $\mathbb{Z}$ is the set of integers. The discrete time quantum walk is defined by using a shift operator and a unitary matrix. Let $\mathbb{C}$ be the set of complex numbers. For $i\in\{1,2,\ldots,d\}$, the shift operator $\tau_i$ is defined by 
\begin{align*}
(\tau_if)(\xvec)=f(\xvec-\evec_{i})\quad (f:\mathbb{Z}^d\longrightarrow\mathbb{C}^{2d}, \ \xvec\in\mathbb{Z}^d),
\end{align*} 
where $\{\evec_1,\evec_2,\ldots,\evec_d\}$ denotes the standard basis of $\mathbb{Z}^d$. Let $A=(a_{ij})_{i,j=1,2,\ldots,2d}$ be a $2d\times 2d$ unitary matrix. We call this unitary matrix the coin matrix. To describe the time evolution of the quantum walk, decompose the unitary matrix $A$ as
\begin{align*}
A=\sum_{i=1}^{2d}P_{i}A,
\end{align*}
where $P_i$ denotes the orthogonal projection onto the one-dimensional subspace $\mathbb{C}\eta_i$ in $\mathbb{C}^{2d}$. Here $\{\eta_1,\eta_2,\ldots,\eta_{2d}\}$ denotes the standard basis on $\mathbb{C}^{2d}$. 
The discrete time quantum walk associated with the coin matrix $A$ is given by
\begin{equation}\label{unitaryop1}
U_A=\sum_{i=1}^d\Big(P_{2i-1}A\tau_{i}^{-1}+P_{2i}A\tau_{i}\Big).
\end{equation}
The state at time $n$ and location $x$ can be expressed by a $2d$-dimensional vector:
\[\Psi_{n}(\xvec)=\begin{bmatrix}\Psi^{1}_{n}(\xvec)\\ \Psi^{2}_{n}(\xvec)\\\vdots\\ \Psi^{2d}_{n}(\xvec) \end{bmatrix}\in\mathbb{C}^{2d}\quad (\xvec\in\mathbb{Z}^d, \ n\in\mathbb{Z}_{\geq}),\]
where $\mathbb{Z}_{\geq}=\{0,1,2,\ldots\}$. For $\Psi_n:\mathbb{Z}^d\longrightarrow\mathbb{C}^{2d}$ and $n\in\mathbb{Z}_{\geq}$, we can rewrite Eq. \eqref{unitaryop1} as
\begin{align}\label{reunitaryop1}
\Psi_{n+1}(\xvec)\equiv(U_A\Psi_{n})(\xvec)=\sum_{i=1}^{d}\Big(P_{2i-1}A\Psi_{n}(\xvec+\evec_i)+P_{2i}A\Psi_{n}(\xvec-\evec_i)\Big).
\end{align} 
This equation means that the particle moves at each step one unit to the $x_i$-axis direction with matrix $P_{2i}A$ or one unit to the $-x_i$-axis direction with matrix $P_{2i-1}A$. For time $n\in\mathbb{Z}_{\geq}$ and location $x\in\mathbb{Z}^d$, we define the measure $\mu_n(\xvec)$ by
$$\mu_n(\xvec)=\|\Psi_n(\xvec)\|_{\mathbb{C}^{2d}}^2,$$
where $\|\cdot\|_{\mathbb{C}^{2d}}$ denotes the standard norm on $\mathbb{C}^{2d}$. Let $\mathbb{R}_{\geq}=[0,\infty)$.  Here we introduce a map $\phi:(\mathbb{C}^{2d})^{\mathbb{Z}^d}\longrightarrow (\mathbb{R}_{\geq})^{\mathbb{Z}^d}$ such that if $\Psi_n:\mathbb{Z}^d\longrightarrow\mathbb{C}^{2d}$ and $\xvec\in\mathbb{Z}^{d}$, 
then $\phi(\Psi_n)(\xvec)=\sum_{j=1}^{2d}|\Psi_n^j(\xvec)|^2\in\mathbb{R}_{\geq}$.
Thus we get
\begin{align*}
\phi(\Psi_n)(\xvec)=\sum_{j=1}^{2d}|\Psi_n^{j}(\xvec)|^2=\mu_n(\xvec),
\end{align*}
namely this map $\phi$ has a role to transform from amplitudes to measures.

To obtain stationary measures of quantum walks on $\mathbb{Z}^d$ defined by the unitary operator $U_A$ defined in Eq. \eqref{unitaryop1}, we use a method of the Fourier transform  introduced in \cite{gjs}. 
Let $f:[-\pi,\pi)^d\longrightarrow\mathbb{C}^{2d}$ and $\textbf{k}=(k_1,k_2,\ldots,k_{d})\in[-\pi,\pi)^d$. The Fourier transform of the function $f$ is defined by the integral
\begin{align*}
(\mathcal{F}f)(\xvec)=\frac{1}{(2\pi)^d}\int_{[-\pi,\pi)^d}e^{i\langle \xvec,  \textbf{k} \rangle}f(\textbf{k})\ d\textbf{k}\qquad(\xvec\in\mathbb{Z}^d),
\end{align*}
where $\langle \xvec,  \textbf{k} \rangle$ is the canonical inner product of $\mathbb{R}^d$. 
Then the inverse of the Fourier transform $\mathcal{F}^*$ is given by 
\begin{align*}
\hat{g}(\textbf{k})\equiv(\mathcal{F}^*g)(\textbf{k})=\sum_{\xvec\in\mathbb{Z}^d}e^{-i\langle \xvec,  \textbf{k} \rangle}\ g(\xvec)\quad\left(g:\mathbb{Z}^d\longrightarrow\mathbb{C}^{2d},\ \textbf{k}\in [-\pi,\pi)^d\right).
\end{align*}
From the inverse of the Fourier transform and Eq. \eqref{reunitaryop1}, we have
\begin{align*}
\hat{\Psi}_{n+1}(\textbf{k})=\hat{U}_A(\textbf{k})\hat{\Psi}_n(\textbf{k}),
\end{align*}
where $\Psi_n:\mathbb{Z}^d\longrightarrow\mathbb{C}^{2d}$ and matrix $\hat{U}_A(\textbf{k})$ is determined by
\begin{align*}                          
\hat{U}_A(\textbf{k})=\sum_{j=1}^{d}\Big(e^{ik_j}P_{2j-1}A+e^{-ik_j}P_{2j}A\Big). 
\end{align*}
We remark that matrix $\hat{U}_A(\textbf{k})$ is a unitary matrix. 
\section{Stationary amplitudes of the Grover walk on $\mathbb{Z}^d$}\label{zdstationary1}
\noindent In this section, we give the definition of the stationary measure for the quantum walk. We define a set of measures, $\mathcal{M}_s(U_A)$, by
\begin{equation*}
\begin{split}
&\mathcal{M}_s(U_A)=\Big\{\mu\in[0,\infty)^{\mathbb{Z}^d}\setminus\{\textbf{0}\};\ there\ exists\ \Psi_0\in\left(\mathbb{C}^D\right)^{\mathbb{Z}^d}\ such\ that\\
&\hspace{6.0cm}\phi(U_A^n\Psi_0)=\mu\ (n=0,1,2,\ldots)\Big\},
\end{split}
\end{equation*}
where $\textbf{0}$ is the zero vector and $D$ is a positive integer. Here $U_A$ is the time evolution operator of quantum walk associated with a unitary matrix $A$. We call this measure $\mu\in\mathcal{M}_s(U_A)$ the stationary measure for the quantum walk defined by the unitary operator $U_A$. If $\mu\in\mathcal{M}_s(U_A)$, then $\mu_n=\mu$ for $n\in\mathbb{Z}_{\geq}$, where $\mu_n$ is the measure of quantum walk given by $U_A$ at time $n$.

Next we consider the following eigenvalue problem of the quantum walk determined by $U_A$:
 \begin{equation}\label{eig.pro}
U_A\Psi=\lambda\Psi\quad(\lambda\in\mathbb{C},\ |\lambda|=1).
\end{equation}
We see that $\phi(\Psi)\in\mathcal{M}_s(U_A)$. Let $D=2d$. Our purpose of this paper is to find stationary measures for our $2d$-state quantum walks on $\mathbb{Z}^d$ by using Eq. \eqref{eig.pro}.
From now on, we treat the stationary amplitude and the stationary measures of the Grover walk on $\mathbb{Z}^d$, where if the function $\Psi$ is satisfied with $\lambda=1$ in Eq. \eqref{eig.pro}, $\Psi$ is called the {\it stationary amplitude}.
 Here, a quantum walk defined by the following $D\times D$ unitary matrix $G=(g_{ij})_{i,j=1,2,\ldots,D}$ is called the Grover walk:
\begin{align}\label{grovermat}
g_{ij}=\frac{2}{D}-\delta_{ij}.
\end{align}
We put $\textbf{k}=(k_1,k_2,\ldots,k_d)$, where $k_j\in[-\pi,\pi)$ $(j=1,2,\ldots,d)$. The Grover walk on $\mathbb{Z}^d$ defined Eqs. \eqref{unitaryop1} and \eqref{grovermat} has an eigenvalue $1$. Furthermore, the unitary matrix $\hat{U}_{G}(\textbf{k})$ induced by the coin matrix $G$ has also an    eigenvalue $1$. We consider an eigenfunction of eigenvalue $1$ of $\hat{U}_{G}(\textbf{k})$. Let $X_j=e^{ik_j}$ and $\overline{X_j}=e^{-ik_j}$ $(j=1,\ldots,d)$.
\begin{lemm}\label{eigenfunction1}
Let $\hat{\Psi}(\textbf{k})$ be an eigenfunction of eigenvalue $1$ of $\hat{U}_{G}(\textbf{k})$.  Then we have
\begin{align}\label{lemma1}
\hat{\Psi}(\textbf{k})=\begin{bmatrix}
(1+X_1)\prod_{l=2}^{d}(1+\overline{X_l})(1+X_l)\\
(1+\overline{X_1})\prod_{l=2}^{d}(1+\overline{X_l})(1+X_l)\\
\vdots\\
(1+X_{j})\prod_{l\ne j}(1+\overline{X_l})(1+X_l)\\
(1+\overline{X_{j}})\prod_{l\ne j}(1+\overline{X_l})(1+X_l)\\
\vdots\\
(1+X_{d})\prod_{l=1}^{d-1}(1+\overline{X_l})(1+X_l)\\
(1+\overline{X_{d}})\prod_{l=1}^{d-1}(1+\overline{X_l})(1+X_l)
\end{bmatrix}.
\end{align}
\end{lemm}
{\bf Proof.} Suppose that $\hat{\Psi}(\textbf{k})$ is an eigenfunction of eigenvalue $1$ of $\hat{U}_{G}(\textbf{k})$. We put $\hat{\Psi}(\textbf{k})={}^T[\hat{\Psi}^1(\textbf{k}),\ldots,\hat{\Psi}^{2d}(\textbf{k})]\in\mathbb{C}^{2d}$, where $T$ means the transposed operation. Then $\hat{U}_{G}(\textbf{k})\hat{\Psi}(\textbf{k})=\hat{\Psi}(\textbf{k})$ is written as
\begin{align*}
       &0=\Big\{(1-d)X_j-d\Big\}\hat{\Psi}^{2j-1}(\textbf{k})+X_j\sum_{l\ne 2j-1}\hat{\Psi}^{l}(\textbf{k}),
    \\
     &0=\Big\{(1-d)\overline{X_j}-d\Big\}\hat{\Psi}^{2j}(\textbf{k})+\overline{X_j}\sum_{l\ne 2j}\hat{\Psi}^{l}(\textbf{k})\qquad(j=1,\ldots,d).                                                                                   
\end{align*}
Thus we get
\begin{align*}
\hat{\Psi}^{2j-1}(\textbf{k})=\frac{1+\overline{X_1}}{1+\overline{X_j}}\hat{\Psi}^{1}(\textbf{k}),\qquad \hat{\Psi}^{2j}(\textbf{k})=\frac{1+\overline{X_1}}{1+X_j}\hat{\Psi}^{1}(\textbf{k}).
\end{align*}
If we take $\hat{\Psi}^{1}(\textbf{k})=(1+X_1)\prod_{l=2}^{d}(1+\overline{X_l})(1+X_l)$, then we have 
\begin{align*}
&\hat{\Psi}^{2j-1}(\textbf{k})=(1+X_{j})\prod_{l\ne j}(1+\overline{X_l})(1+X_l),\\
&\hat{\Psi}^{2j}(\textbf{k})=(1+\overline{X_{j}})\prod_{l\ne j}(1+\overline{X_l})(1+X_l).
\end{align*}
The proof of Lemma \ref{eigenfunction1} is complete.
\vspace{0.8cm}

\noindent For $\uvec=(u_1,\ldots,u_d)\in\mathbb{Z}^d$, we put a subset $K_{\uvec}^d\subset\mathbb{Z}^d$ as
\begin{align*}
K_{\uvec}^d=\left\{\xvec=(x_1,x_2,\ldots,x_{d})\in\mathbb{Z}^d;\sqrt{\sum_{i=1}^{d}(x_i-u_i)^2}\leq\sqrt{d}\right\}.
\end{align*}
\noindent Remark that ${}^{\#}K_{\uvec}^d=3^d$, where ${}^{\#}X$ means the number of elements in a set $X$.  Let $\hat{\Psi}$ be a map $\hat{\Psi}:[-\pi,\pi)^d\longrightarrow\mathbb{C}^{2d}$ given by Eq. \eqref{lemma1}. 
 For $\xvec\in\mathbb{Z}^{d}$, we define 
\begin{align*}
\Psi_s^{(\textbf{0})}(\xvec)\equiv(\mathcal{F}\hat{\Psi})(\xvec), 
\end{align*}
where $\textbf{0}=(0,\ldots,0)\in\mathbb{Z}^d$ is the origin of $\mathbb{Z}^d$. Each $\cvec\in K_{\textbf{0}}^d$, we set the $2d$-dimensional complex vector $\avec_{\cvec}$:
\begin{align*}
\avec_{\cvec}\equiv\Psi_s^{(\textbf{0})}(\cvec).
\end{align*}
\begin{rem}
{\it If a unitary matrix $A$ is given, the vector $\avec_{\cvec}\in\mathbb{C}^{2d}$ is uniquely determined}.
\end{rem}
For $\uvec\in\mathbb{Z}^d$, we define the function $\Psi_s^{(\uvec)}:\mathbb{Z}^d\longrightarrow\mathbb{C}^{2d}$ as 
\begin{align}\label{cdeffunc}
\Psi_s^{(\uvec)}(\xvec)=\sum_{\cvec\in K_{\textbf{0}}^d}\avec_{\cvec}\ \delta_{\uvec+\cvec}(\xvec).
\end{align}
Equation \eqref{cdeffunc} means that the particle stays in the $3^d$-neighboring points centered at the point $(u_1,\ldots,u_d)\in\mathbb{Z}^d$ with the weight $\avec_{\cvec}$ at the point $\cvec\in K_{\textbf{0}}^d$. Thus ${\rm supp}\ \left(\Psi_s^{(\uvec)}\right)$ is a finite support of $\mathbb{Z}^d$, namely
\begin{align*}
{\rm supp}\ \left(\Psi_s^{(\uvec)}\right)=K_{\uvec}^d,
\end{align*}
where ${\rm supp}\ (f)$ is defined by 
\begin{align*}
{\rm supp}\ (f)=\overline{\Big\{\xvec\in\mathbb{Z}^d;f(\xvec)\ne0\Big\}}.
\end{align*}
\begin{thm}\label{zdstationary2}
We consider the stationary measures of the Grover walk defined Eq. \eqref{grovermat} on $\mathbb{Z}^d$. Let $\{\varphi_{\uvec}\}_{\uvec\in\mathbb{Z}^d}$ be a sequence of complex numbers except for $\varphi\equiv0$. Here $\varphi\equiv0$ means that $\varphi_{\uvec}=0$ $(\uvec\in\mathbb{Z}^d)$. We define the function $\Psi_s^{(\varphi)}(\xvec)$ as 
\begin{align*}
\Psi_{s}^{(\varphi)}(\xvec)=\sum_{\uvec\in\mathbb{Z}^d}\varphi_{\uvec}\ \Psi_s^{(\uvec)}(\xvec).
\end{align*}
Then $\Psi_s^{(\varphi)}(\xvec)$ is  an eigenfunction of eigenvalue $1$ of $U_G$. As a corollary, we obtain 
\begin{align*}
\phi(\Psi_s^{(\varphi)})\in\mathcal{M}_s(U_G).
\end{align*}
That is to say, this $\Psi_s^{(\varphi)}(\xvec)$ gives us the stationary measures of the Grover walk on $\mathbb{Z}^d$.
\end{thm}
{\bf Proof.} 
We put $\hat{\Psi}_n(\textbf{k})=(\hat{U}_G)^n\hat{\Psi}(\textbf{k})$, where $\hat{\Psi}(\textbf{k})$ is given by Eq. \eqref{lemma1}. Since $(\hat{U}_G)^n\hat{\Psi}(\textbf{k})=\hat{\Psi}(\textbf{k})$, it holds $(\mathcal{F}\hat{\Psi}_n)(\xvec)=(\mathcal{F}\hat{\Psi})(\xvec)$ for $\xvec\in\mathbb{Z}^d$. Thus we get 
\begin{align*}
U_G^n\ \Psi_s^{(\textbf{0})}(\xvec)=\Psi_s^{(\textbf{0})}(\xvec)\qquad(\xvec\in\mathbb{Z}^d).
\end{align*}
Moreover, we see that 
\begin{align*}
U_G^n\ \Psi_s^{(\textbf{\uvec})}(\xvec)=\Psi_s^{(\textbf{\uvec})}(\xvec)\qquad(\uvec\in\mathbb{Z}^d).
\end{align*}
Then this $\Psi_s^{(\uvec)}(\xvec)$ is a stationary amplitude. Let $\{\varphi_{\uvec}\}_{\uvec\in\mathbb{Z}^d}$ be a sequence of complex numbers except for $\varphi\equiv0$. Here $\varphi\equiv0$ means that $\varphi_{\uvec}=0$ $(\uvec\in\mathbb{Z}^d)$. We define the function $\Psi_s^{(\varphi)}(\xvec)$ as 
\begin{align*}
\Psi_s^{(\varphi)}(\xvec)=\sum_{\uvec\in\mathbb{Z}^d}\varphi_{\uvec}\ \Psi_s^{(\uvec)}(\xvec).
\end{align*}
For $n\in\mathbb{Z}_{\geq}$, we have
\begin{equation}\label{stationaryamp1}
\begin{split}
U_G^n\ \Psi_s^{(\varphi)}&=U_G^n\ \Big(\sum_{\uvec\in\mathbb{Z}^d}\varphi_{\uvec}\ \Psi_s^{(\uvec)}\Big)=\sum_{\uvec\in\mathbb{Z}^d}\varphi_{\uvec}\Big(U_G^n\ \Psi_s^{(\uvec)}\Big)\\
&=\sum_{\uvec\in\mathbb{Z}^d}\varphi_{\uvec}\ \Psi_s^{(\uvec)}=\Psi_s^{(\varphi)}.
\end{split}
\end{equation}
Then $\Psi_s^{(\varphi)}(\xvec)$ gives us the stationary amplitude of the Grover walk on $\mathbb{Z}^d$. As a corollary, we obtain $\phi\big(\Psi_s^{(\varphi)}\big)\in\mathcal{M}_s(U_G)$ from Eq. \eqref{stationaryamp1}. This completes the proof of Theorem \ref{zdstationary2}.
\section{Examples}\label{examples}
In this section, we present two examples: one is a class of quantum walks including the Grover walk  on $\mathbb{Z}^2$ and the other is the Grover walk on $\mathbb{Z}^3$.
\begin{exa}
\end{exa}
We consider a class of four-state quantum walks on $\mathbb{Z}^2$ including the Grover walk determined by $4\times4$ unitary matrices $A_1$ introduced by Watabe et al. \cite{wkkk} as
\begin{align*}
A_1=
\begin{bmatrix}
-p&q&\sqrt{pq}&\sqrt{pq}\\
q&-p&\sqrt{pq}&\sqrt{pq}\\
\sqrt{pq}&\sqrt{pq}&-q&p\\
\sqrt{pq}&\sqrt{pq}&p&-q
\end{bmatrix},
\end{align*}
where $q=1-p$ and $p\in(0,1)$. Note that the quantum walk given by $p=q=\frac{1}{2}$ is the Grover walk. Let $X_l=e^{ik_{l}}$ $(l=1,2)$. Thus the unitary matrix $\hat{U}_{A_1}(\textbf{k})$ becomes
\begin{align*}
\hat{U}_{A_1}(\textbf{k})=
\begin{bmatrix}
-pX_1&qX_1&\sqrt{pq}X_1&\sqrt{pq}X_1\\
q\overline{X_1}&-p\overline{X_1}&\sqrt{pq}\overline{X_1}&\sqrt{pq}\overline{X_1}\\
\sqrt{pq}X_2&\sqrt{pq}X_2&-qX_2&pX_2\\
\sqrt{pq}\overline{X_2}&\sqrt{pq}\overline{X_2}&p\overline{X_2}&-q\overline{X_2}
\end{bmatrix}.
\end{align*}
We focus on an eigenfunction $\hat{\Psi}(\textbf{k})$ of eigenvalue $1$ of $\hat{U}_{A_1}(\textbf{k})$. 
\begin{align*}
\begin{split}
\hat{\Psi}(\textbf{k})&=\begin{bmatrix}
(1+X_1)(1+\overline{X_2})(1+X_2)\\
(1+\overline{X_1})(1+\overline{X_2})(1+X_2)\\
\frac{\sqrt{pq}}{q}(1+X_2)(1+\overline{X_1})(1+X_1)\\
\frac{\sqrt{pq}}{q}(1+\overline{X_2})(1+\overline{X_1})(1+X_1)
\end{bmatrix}\\
&=\begin{bmatrix}
(2+2X_1+X_2+\overline{X_2}+X_1\overline{X_2}+X_1X_2)\\
(2+2\overline{X_1}+X_2+\overline{X_2}+\overline{X_1}X_2+\overline{X_1X_2})\\
\frac{\sqrt{pq}}{q}(2+2X_2+X_1+\overline{X_1}+\overline{X_1}X_2+X_1X_2)\\
\frac{\sqrt{pq}}{q}(2+2\overline{X_2}+X_1+\overline{X_1}+X_1\overline{X_2}+\overline{X_1X_2})
\end{bmatrix}.
\end{split}
\end{align*}
Since $\avec\equiv\Psi_s^{(\textbf{0})}=\mathcal{F}\hat{\Psi}$, we get 
\begin{align*}
\begin{split}
\avec&=
\begin{bmatrix}
2\\
2\\
2\frac{\sqrt{pq}}{q}\\
2\frac{\sqrt{pq}}{q}
\end{bmatrix}\delta_{(0,0)}
+
\begin{bmatrix}
1\\
1\\
0\\
2\frac{\sqrt{pq}}{q}
\end{bmatrix}\delta_{(0,1)}
+
\begin{bmatrix}
0\\
2\\
\frac{\sqrt{pq}}{q}\\
\frac{\sqrt{pq}}{q}
\end{bmatrix}\delta_{(1,0)}
+
\begin{bmatrix}
1\\
1\\
2\frac{\sqrt{pq}}{q}\\
0
\end{bmatrix}\delta_{(0,-1)}\\
&+
\begin{bmatrix}
2\\
0\\
\frac{\sqrt{pq}}{q}\\
\frac{\sqrt{pq}}{q}
\end{bmatrix}\delta_{(-1,0)}
+
\begin{bmatrix}
0\\
1\\
0\\
\frac{\sqrt{pq}}{q}
\end{bmatrix}\delta_{(1,1)}
+
\begin{bmatrix}
0\\
1\\
\frac{\sqrt{pq}}{q}\\
0
\end{bmatrix}\delta_{(1,-1)}\\
&+
\begin{bmatrix}
1\\
0\\
\frac{\sqrt{pq}}{q}\\
0
\end{bmatrix}\delta_{(-1,-1)}
+
\begin{bmatrix}
1\\
0\\
0\\
\frac{\sqrt{pq}}{q}
\end{bmatrix}\delta_{(-1,1)}.
\end{split}
\end{align*}
For $\uvec=(u_1,u_2)\in\mathbb{Z}^2$, we put the function $\Psi_s^{(\uvec)}:\mathbb{Z}^2\longrightarrow\mathbb{C}^{4}$ as 
\begin{align*}
\Psi_s^{(\uvec)}(\xvec)=\sum_{\cvec\in K_{\textbf{0}}^2}\avec_{\cvec}\ \delta_{\uvec+\cvec}(\xvec).
\end{align*}
Thus we obtain
\begin{align*}
{\rm supp}\ \left(\Psi_s^{(u_1,u_2)}\right)&=
\Big\{(u_1,u_2),\ (u_1\pm1,u_2),\ (u_1,u_2\pm1),\ (u_1\pm1,u_2\pm1)\Big\}\\
&=\left\{\xvec=(x_1,x_2)\in\mathbb{Z}^2;\sqrt{\sum_{i=1}^2(x_i-u_i)^2}\leq \sqrt{2}\right\}\subset\mathbb{Z}^2.
\end{align*}
This $\Psi_s^{(u_1,u_2)}(x_1,x_2)$ gives us the stationary amplitude for the Grover walk on $\mathbb{Z}^2$. Let $\{\varphi_{(u_1,u_2)}\}_{(u_1,u_2)\in\mathbb{Z}^2}$ be a sequence of complex numbers except for $\varphi\equiv0$. We define the following stationary amplitude:
\begin{align*}
\Psi_s^{(\varphi)}(x_1,x_2)=\sum_{(u_1,u_2)\in\mathbb{Z}^2}\varphi_{(u_1,u_2)}\  
\Psi_s^{(u_1,u_2)}(x_1,x_2).
\end{align*}
If we take $\Psi_s^{(\varphi)}=\varphi_{\textbf{0}}\ \Psi_s^{(\textbf{0})}$, $p=q=\frac{1}{2}$ and $\varphi_{\textbf{0}}=\frac{1}{4\sqrt{3}}$ $\big(\varphi_{\uvec}=0\ (\uvec\ne\textbf{0})\big)$, we have
\begin{equation*}
\begin{split}
\mu(x_1,x_2)&=\|\Psi_s^{(\varphi)}(x_1,x_2)\|_{\mathbb{C}^4}\\
&=\Big(\frac{1}{3}\delta_{(0,0)}+\frac{1}{8}\big(\delta_{(0,1)}+\delta_{(0,-1)}+\delta_{(1,0)}+\delta_{(-1,0)}\big)\\
&+\frac{1}{24}\big(\delta_{(1,-1)}+\delta_{(-1,1)}+\delta_{(1,1)}+\delta_{(-1,-1)}\big)\Big)(x_1,x_2).
\end{split}
\end{equation*}
We remark that this function $\Psi_s^{(\varphi)}$ belongs to the Hilbert space $\ell^2(\mathbb{Z}^2,\mathbb{C}^4)$ and this measure $\mu$ is a probability measure.
\begin{exa}
\end{exa}
We consider the Grover walk on $\mathbb{Z}^3$ determined by $6\times6$ unitary matrix $G$ defined in  Eq. \eqref{grovermat}. Let $X_l=e^{ik_{l}}$ $(l=1,2,3)$. Thus the unitary matrix $\hat{U}_{G}(\textbf{k})$ becomes
\begin{align*}
\hat{U}_{G}(\textbf{k})=\frac{1}{3}
\begin{bmatrix}
-2X_1&X_1&X_1&X_1&X_1&X_1\\
\overline{X_1}&-2\overline{X_1}&\overline{X_1}&\overline{X_1}&\overline{X_1}&\overline{X_1}\\
X_2&X_2&-2X_2&X_2&X_2&X_2\\
\overline{X_2}&\overline{X_2}&\overline{X_2}&-2\overline{X_2}&\overline{X_2}&\overline{X_2}\\
X_3&X_3&X_3&X_3&-2X_3&X_3\\
\overline{X_3}&\overline{X_3}&\overline{X_3}&\overline{X_3}&\overline{X_3}&-2\overline{X_3}
\end{bmatrix}.
\end{align*}
We focus on an eigenfunction $\hat{\Psi}(\textbf{k})$ of eigenvalue $1$ of $\hat{U}_{G}(\textbf{k})$. Then we get
\begin{align*}
\hat{\Psi}(\textbf{k})={}^T\Big[\hat{\Psi}^1(\textbf{k}),\ \hat{\Psi}^2(\textbf{k}),\ 
\hat{\Psi}^3(\textbf{k}),\ 
\hat{\Psi}^4(\textbf{k}),\ 
\hat{\Psi}^5(\textbf{k}),\ 
\hat{\Psi}^6(\textbf{k})\Big]\in\mathbb{C}^{6},
\end{align*}
where 
\begin{equation*}
\begin{split}
\hat{\Psi}^1(\textbf{k})&=2^2+2(X_2+\overline{X_2}+X_3+\overline{X_3})+
2^2X_1+2X_1(X_2+\overline{X_2}+X_3+\overline{X_3}) \\
&+X_2X_3+X_2\overline{X_3}+\overline{X_2}X_3+\overline{X_2X_3}
+X_1(X_2X_3+X_2\overline{X_3}+\overline{X_2}X_3+\overline{X_2X_3}),\\
\hat{\Psi}^2(\textbf{k})&=2^2+2(X_2+\overline{X_2}+X_3+\overline{X_3})+2^2\overline{X_1}+2\overline{X_1}(X_2+\overline{X_2}+X_3+\overline{X_3}) \\
&+X_2X_3+X_2\overline{X_3}+\overline{X_2}X_3+\overline{X_2X_3}
+\overline{X_1}(X_2X_3+X_2\overline{X_3}+\overline{X_2}X_3+\overline{X_2X_3}),\\
\hat{\Psi}^3(\textbf{k})&=2^2+2(X_1+\overline{X_1}+X_3+\overline{X_3})+2^2X_2+2X_2(X_1+\overline{X_1}+X_3+\overline{X_3}) \\
&+X_1X_3+X_1\overline{X_3}+\overline{X_1}X_3+\overline{X_1X_3}
+X_2(X_1X_3+X_1\overline{X_3}+\overline{X_1}X_3+\overline{X_1X_3}),\\
\hat{\Psi}^4(\textbf{k})&=2^2+2(X_1+\overline{X_1}+X_3+\overline{X_3})+2^2\overline{X_2}+2\overline{X_2}(X_1+\overline{X_1}+X_3+\overline{X_3}) \\
&+X_1X_3+X_1\overline{X_3}+\overline{X_1}X_3+\overline{X_1X_3}
+\overline{X_2}(X_1X_3+X_1\overline{X_3}+\overline{X_1}X_3+\overline{X_1X_3}),\\
\hat{\Psi}^5(\textbf{k})&=2^2+2(X_1+\overline{X_1}+X_2+\overline{X_2})+2^2X_3+2X_3(X_1+\overline{X_1}+X_2+\overline{X_2}) \\
&+X_1X_2+X_1\overline{X_2}+\overline{X_1}X_2+\overline{X_1X_2}
+X_3(X_1X_2+X_1\overline{X_2}+\overline{X_1}X_2+\overline{X_1X_2}),\\
\hat{\Psi}^6(\textbf{k})&=2^2+2(X_1+\overline{X_1}+X_2+\overline{X_2})+2^2\overline{X_3}+2\overline{X_3}(X_1+\overline{X_1}+X_2+\overline{X_2}) \\
&+X_1X_2+X_1\overline{X_2}+\overline{X_1}X_2+\overline{X_1X_2}
+\overline{X_3}(X_1X_2+X_1\overline{X_2}+\overline{X_1}X_2+\overline{X_1X_2}).
\end{split}
\end{equation*}
Since $\avec\equiv\Psi_s^{(\textbf{0})}=\mathcal{F}\hat{\Psi}$, we have
\begin{align*}
\begin{split}
\avec&=
\begin{bmatrix}
4\\
4\\
4\\
4\\
4\\
4
\end{bmatrix}\delta_{(0,0,0)}
+\begin{bmatrix}
2\\
2\\
0\\
4\\
2\\
2
\end{bmatrix}\delta_{(0,1,0)}
+\begin{bmatrix}
0\\
4\\
2\\
2\\
2\\
2
\end{bmatrix}\delta_{(1,0,0)}
+\begin{bmatrix}
2\\
2\\
4\\
0\\
2\\
2
\end{bmatrix}\delta_{(0,-1,0)}\\
&+\begin{bmatrix}
4\\
0\\
2\\
2\\
2\\
2
\end{bmatrix}\delta_{(-1,0,0)}
+\begin{bmatrix}
0\\
2\\
0\\
2\\
1\\
1
\end{bmatrix}\delta_{(1,1,0)}
+\begin{bmatrix}
0\\
2\\
2\\
0\\
1\\
1
\end{bmatrix}\delta_{(1,-1,0)}
+\begin{bmatrix}
2\\
0\\
2\\
0\\
1\\
1
\end{bmatrix}\delta_{(-1,-1,0)}\\
&+\begin{bmatrix}
2\\
0\\
0\\
2\\
1\\
1
\end{bmatrix}\delta_{(-1,1,0)} 
+\begin{bmatrix}
2\\
2\\
2\\
2\\
4\\
0
\end{bmatrix}\delta_{(0,0,-1)}
+\begin{bmatrix}
1\\
1\\
0\\
2\\
2\\
0
\end{bmatrix}\delta_{(0,1,-1)}
+\begin{bmatrix}
0\\
2\\
1\\
1\\
2\\
0
\end{bmatrix}\delta_{(1,0,-1)}\\
&+\begin{bmatrix}
1\\
1\\
2\\
0\\
2\\
0
\end{bmatrix}\delta_{(0,-1,-1)}
+\begin{bmatrix}
2\\
0\\
1\\
1\\
2\\
0
\end{bmatrix}\delta_{(-1,0,-1)}
+\begin{bmatrix}
0\\
1\\
0\\
1\\
1\\
0
\end{bmatrix}\delta_{(1,1,-1)}
+\begin{bmatrix}
0\\
1\\
1\\
0\\
1\\
0
\end{bmatrix}\delta_{(1,-1,-1)}\\ 
&+\begin{bmatrix} 
1\\
0\\
1\\
0\\
1\\
0
\end{bmatrix}\delta_{(-1,-1,-1)}
+\begin{bmatrix}
1\\
0\\
0\\
1\\
1\\
0
\end{bmatrix}\delta_{(-1,1,-1)}
+\begin{bmatrix}
2\\
2\\
2\\
2\\
0\\
4
\end{bmatrix}\delta_{(0,0,1)}
+\begin{bmatrix}
1\\
1\\
0\\
2\\
0\\
2
\end{bmatrix}\delta_{(0,1,1)}\\
&+\begin{bmatrix}
0\\
2\\
1\\
1\\
0\\
2
\end{bmatrix}\delta_{(1,0,1)}
+\begin{bmatrix}
1\\
1\\
2\\
0\\
0\\
2
\end{bmatrix}\delta_{(0,-1,1)}
+\begin{bmatrix}
2\\
0\\
1\\
1\\
0\\
2
\end{bmatrix}\delta_{(-1,0,1)}
+\begin{bmatrix}
0\\
1\\
0\\
1\\
0\\
1
\end{bmatrix}\delta_{(1,1,1)}\\
&+\begin{bmatrix}
0\\
1\\
1\\
0\\
0\\
1
\end{bmatrix}\delta_{(1,-1,1)}
+\begin{bmatrix}
1\\
0\\
1\\
0\\
0\\
1
\end{bmatrix}\delta_{(-1,-1,1)}
+\begin{bmatrix}
1\\
0\\
0\\
1\\
0\\
1
\end{bmatrix}\delta_{(-1,1,1)}
.
\end{split}
\end{align*}
Each $\uvec=(u_1,u_2,u_3)\in\mathbb{Z}^3$, we give the function $\Psi_s^{(\uvec)}:\mathbb{Z}^3\longrightarrow\mathbb{C}^{6}$ as 
\begin{align*}
\Psi_s^{(\uvec)}(\xvec)=\sum_{\cvec\in K_{\textbf{0}}^3}\avec_{\cvec}\ \delta_{\uvec+\cvec}(\xvec).
\end{align*}
Then we have
\begin{equation*}
\begin{split}
{\rm supp}\ \left(\Psi_s^{(u_1,u_2,u_3)}\right)=
\Big\{&(u_1,u_2,u_3),\ (u_1\pm1,u_2\pm1,u_3), \ (u_1,u_2\pm1,u_3),\\  
&(u_1\pm1,u_2,u_3),\ (u_1,u_2,u_3\pm1),\ (u_1\pm1,u_2,u_3\pm1),\\ 
&(u_1,u_2\pm1,u_3\pm1),\ (u_1\pm1,u_2\pm1,u_3\pm1)\Big\}\\
&\hspace{-0.7cm}=\left\{\xvec=(x_1,x_2,x_3)\in\mathbb{Z}^3;\sqrt{\sum_{i=1}^3(x_i-u_i)^2}\leq \sqrt{3}\right\}\subset\mathbb{Z}^3.
\end{split}
\end{equation*}
This $\Psi_s^{(u_1,u_2,u_3)}(x_1,x_2,x_3)$ gives us the stationary amplitude for the Grover walk on $\mathbb{Z}^3$. Let $\{\varphi_{(u_1,u_2,u_3)}\}_{(u_1,u_2,u_3)\in\mathbb{Z}^3}$ be a sequence of complex numbers except for $\varphi\equiv0$. We can take the following stationary amplitude:
\begin{align*}
\Psi_s^{(\varphi)}(x_1,x_2,x_3)=\sum_{(u_1,u_2,u_3)\in\mathbb{Z}^3}\varphi_{(u_1,u_2,u_3)}\ \Psi_s^{(u_1,u_2,u_3)}(x_1,x_2,x_3).
\end{align*}
Thus $\Psi_s^{(\varphi)}(x_1,x_2,x_3)$ gives us the stationary amplitude for the Grover walk on $\mathbb{Z}^3$. If we take $\Psi_s^{(\varphi)}=\varphi_{\textbf{0}}\ \Psi_s^{(\textbf{0})}$ and $\varphi_{\textbf{0}}=\frac{1}{12\sqrt{3}}$ $\big(\varphi_{\uvec}=0\ (\uvec\ne\textbf{0})\big)$, we obtain
\begin{align*}
\mu(x_1,x_2,x_3)&=\|\Psi_s^{(\varphi)}(x_1,x_2,x_3)\|_{\mathbb{C}^6}\\
&=\Big(\frac{2}{9}\delta_{(0,0,0)}+\frac{2}{27}\mu_1+\frac{1}{144}\mu_2+\frac{5}{216}\mu_3\Big)(x_1,x_2,x_3).
\end{align*}
where
\begin{align*}
\begin{split}
&\mu_1=\delta_{(0,0,1)}+\delta_{(0,0,-1)}+\delta_{(0,1,0)}+\delta_{(0,-1,0)}+\delta_{(1,0,0)}+\delta_{(-1,0,0)},\\
&\mu_2=\delta_{(1,1,1)}+\delta_{(-1,-1,-1)}+\delta_{(1,1,-1)}+\delta_{(-1,1,1)}+\delta_{(1,-1,1)}+\delta_{(1,-1,-1)}\\
&\hspace{0.40cm}+\delta_{(-1,-1,1)}+\delta_{(-1,1,-1)},\\
&\mu_3=\delta_{(0,1,1)}+\delta_{(0,-1,-1)}+\delta_{(0,1,-1)}+\delta_{(0,-1,1)}+\delta_{(1,0,1)}+\delta_{(1,0,-1)}\\
&\hspace{0.40cm}+\delta_{(-1,0,-1)}+\delta_{(-1,0,1)}+\delta_{(1,1,0)}+\delta_{(-1,-1,0)}+\delta_{(1,-1,0)}+\delta_{(-1,1,0)}.
\end{split}
\end{align*}
We remark that this function $\Psi_s^{(\varphi)}$ belongs to the Hilbert space $\ell^2(\mathbb{Z}^3,\mathbb{C}^6)$ and this measure $\mu$ is a probability measure.


\section{Conclusion}\label{conclusion}
\noindent
There has never been study on stationary measures of quantum walks on the higher-dimensional lattice. Thus this paper presented the stationary measures of the Grover walk on $\mathbb{Z}^d$. We remark that we can easily extend our result to the shift operator in a more general setting on $\mathbb{Z}^d$. 

In a future work, we would like to investigate the relationship between the stationary measure and the limit measure for quantum walks on $\mathbb{Z}^d$. However, even the explicit form of the limit measure $\displaystyle\mu_{\infty}(\xvec)$ of the Grover walk on $\mathbb{Z}^d\ (d\geq2)$ is not known. Here, we define the limit measure $\mu_{\infty}$ by
\begin{align*}
\mu_{\infty}(\xvec)=\lim_{n\to\infty}\mu_n(\xvec)\quad(\xvec\in\mathbb{Z}^d),
\end{align*}
if the right-hand side exists. So we need to obtain the limit measure of the Grover walk on $\mathbb{Z}^d\ (d\geq2)$. As a related work, Endo et al. \cite{ekk2} clarified the relationship between the stationary measure and the limit measure in the case of the three-state Grover walk on $\mathbb{Z}$.
   
\par
\
\par
\noindent
{\bf Acknowledgements}
\noindent
This work is partially supported by the Grant-in-Aid for Scientific Research (Challenging Exploratory Research) of Japan Society for the Promotion of Science (Grant No.15K13443).

\par
\
\par

\begin{small}
\bibliographystyle{jplain}

\begin{thebibliography}{10}
\bibitem{adz}
Y. Aharonov, L. Davidovich and N. Zagury, {\it Quantum random walks}, Phys. Rev. A \textbf{48}, pp.1687-1690 (1993).
\bibitem{eekst}
S. Endo, T. Endo, N. Konno, E. Segawa and M. Takei, {\it Limit theorems of a two-phase quantum walk with one defect}, Quantum Inf. Comput. \textbf{15}, pp.1373-1396 (2015).
\bibitem{ekk1}
T. Endo, H. Kawai and N. Konno, {\it The stationary measure for diagonal quantum walk with one defect}, arXiv:1603.08948 (2016).
\bibitem{ekk2}
T. Endo, H. Kawai and N. Konno, {\it Stationary measures for the three-state Grover walk with one defect in one dimension}, RIMS Kokyuroku 2010, pp.45-55 (2016).
\bibitem{ek}
T. Endo and N. Konno, {\it The stationary measure of a space-inhomogeneous quantum walk on the line}, Yokohama Math. J., \textbf{60}, pp.33-47 (2014).
\bibitem{eko}
T. Endo, N. Konno and H. Obuse, {\it Relation between two-phase quantum walks and the topological invariant}, arXiv:1511.04230 (2015).
\bibitem{ekst}
T. Endo, N. Konno, E. Segawa and M. Takei, {\it A one-dimensional Hadamard walk with one defect}, Yokohama Math. J., \textbf{60}, pp.49-90 (2014).
\bibitem{fgmb}
C. Di Franco, M. Mc Gettrick, T. Machida and Th. Busch, {\it Alternate two-dimensional quantum walk with a single-qubit coin}, Phys. Rev. A \textbf{84}, 042337 (2011).
\bibitem{gjs}
G. Grimmett, S. Janson and P. F. Scudo, {\it Weak limits for quantum random walks}, Phys. Rev. E \textbf{69}, 026119 (2004).
\bibitem{hkss}
Y. Higuchi, N. Konno, I. Sato and E. Segawa, {\it Spectral and asymptotic properties of Grover walks on crystal lattices}, J. Funct. Anal. \textbf{267}, pp.4197-4235 (2014).
\bibitem{ikk}
N. Inui, Y. Konishi and N. Konno, {\it Localization of two-dimensional quantum walks}, Phys. Rev. A \textbf{69}, 052323 (2004).
\bibitem{iks}
N. Inui, N. Konno and E. Segawa, {\it One-dimensional three-state quantum walk}, Phys. Rev. E \textbf{72}, 056112 (2005).
\bibitem{kkk}
H. Kawai, T. Komatsu and N. Konno, {\it Stationary measures of three-state quantum walks on the one-dimensional lattice}, arXiv:1702.01523 (2017).
\bibitem{k}
T. Komatsu, {\it Limiting distributions of quantum walks on the square lattice}, Yokohama Math. J. \textbf{61}, pp.67-86 (2015).
\bibitem{ko1}
N. Konno, {\it A new type of limit theorems for the one-dimensional quantum random walk}, J. Math. Soc. Japan \textbf{57}, pp.1179-1195 (2005).
\bibitem{ko2}
N. Konno, {\it The uniform measure for discrete-time quantum walks in one dimension}. Quantum Inf. Process.  \textbf{13}, pp.1103-1125 (2014).
\bibitem{kls}
N. Konno, T. Luczak and E. Segawa, {\it Limit measures of inhomogeneous discrete-time quantum walks in one dimension}, Quantum Inf. Process. \textbf{12}, pp.33-53 (2013).
\bibitem{kt}
N. Konno and M. Takei, {\it The non-uniform stationary measure for discrete-time quantum walks in one dimension}, Quantum Inf. Comput. \textbf{15}, pp.1060-1075 (2015).
\bibitem{mckb}
T. Machida, C. M. Chandrashekar, N. Konno and Th. Busch, {\it Limit distributions for different forms of four-state quantum walks on a two-dimensional lattice}, Quantum Inf. Comput. \textbf{15}, pp.1248-1258 (2015).
\bibitem{mw}
K. Manouchehri and J. Wang, {\it Physical Implementation of Quantum Walks}, Springer (2013).
\bibitem{por}
R. Portugal, {\it Quantum Walks and Search Algorithms}, Springer (2013).
\bibitem{tate}
T. Tate, {\it Eigenvalues, absolute continuity and localizations for periodic unitary transition operators}, arXiv:1411.4215 (2014).
\bibitem{wlw}
C. Wang, X. Lu and W. Wang, {\it The stationary measure of a space-inhomogeneous three-state quantum walk on the line}, Quantum Inf. Process. \textbf{14}, pp.867-880 (2015).
\bibitem{wkkk}
K. Watabe, N. Kobayashi, M. Katori and N. Konno, {\it Limit distributions of two-dimensional quantum walks}, Phys. Rev. A \textbf{77}, 062331 (2008).




















\end{thebibliography}

\end{small}

\end{document}